\documentclass[preprint,amsmath,superscriptaddress,prl]{revtex4}  
\usepackage[dvips]{graphicx}
\usepackage{epsfig}
\usepackage{color}
\usepackage{subfigure}
\usepackage{soul}

\begin{document} 

\title{
The brevity law as a scaling law, \\
and a possible origin of Zipf's law
for word frequencies
}
\author{\'Alvaro Corral}
\affiliation{%
Centre de Recerca Matem\`atica,
Edifici C, Campus Bellaterra,
E-08193 Barcelona, Spain
}\affiliation{Departament de Matem\`atiques,
Facultat de Ci\`encies,
Universitat Aut\`onoma de Barcelona,
E-08193 Barcelona, Spain
}\affiliation{Barcelona Graduate School of Mathematics, 
Edifici C, Campus Bellaterra,
E-08193 Barcelona, Spain
}\affiliation{Complexity Science Hub Vienna,
Josefst\"adter Stra$\beta$e 39,
1080 Vienna,
Austria
}
\author{Isabel Serra}
\affiliation{%
Centre de Recerca Matem\`atica,
Edifici C, Campus Bellaterra,
E-08193 Barcelona, Spain
}\affiliation{%
Barcelona Supercomputing Center - Centro Nacional de Supercomputaci\'on,
Jordi Girona 29, E-08034 Barcelona, Spain
}

\begin{abstract} 
An important body of quantitative linguistics is constituted by a series
of statistical laws about language usage.
Despite the importance of these linguistic laws, 
some of them are poorly formulated, and,
more importantly,
there is no unified framework that encompasses all them.
This paper presents a new perspective to establish a connection between 
different statistical linguistic laws.
Characterizing each word type by two random variables,
length (in number of characters) and absolute frequency,
we show that the corresponding 
bivariate joint probability distribution shows a rich and precise phenomenology,
with the type-length and the type-frequency distributions as its two marginals, 
and the conditional distribution of frequency at fixed length
providing a clear formulation for the brevity-frequency phenomenon.
The type-length distribution turns out to be well fitted by a gamma distribution
(much better than with the previously proposed lognormal),
and the conditional frequency distributions at fixed length
display power-law-decay behavior with a fixed exponent $\alpha\simeq 1.4$ and a characteristic-frequency crossover that scales as an inverse power 
$\delta\simeq 2.8$ of length, which implies the fulfilment of a scaling law
analogous to those found in the thermodynamics of critical phenomena.
As a by-product, we find a possible model-free explanation for the origin of Zipf's law, which should arise as a mixture of conditional frequency distributions governed
by the crossover length-dependent frequency.
%
%
%
%
%
%
%
\end{abstract} 
\maketitle


\section{Introduction}

The usage of language, both in its written and oral forms, follows very clear statistical regularities \cite{Hdez_FCancho_book}. 
One of the goals of quantitative linguistics is to unveil, analyze, explain, and exploit such linguistic statistical laws. 
Perhaps the clearest example of a statistical law in language usage is Zipf's law, which quantifies the frequency of occurrence of words in texts and speech 
\cite{Zipf_1949,Baayen,Baroni2009,Zanette_book,Piantadosi,Moreno_Sanchez}, 
establishing that there is no unarbitrary way to distinguish between rare and common words. 
Surprisingly, Zipf's law is not only a linguistic law, but seems to be a rather common phenomenon in complex systems where discrete units self-organize into groups, or types 
(persons into cities, money into persons, etc. \cite{Corral_Cancho}). 
Another well-known linguistic pattern, related to Zipf's law, 
is Herdan's law, also called Heaps' law
\cite{Mandelbrot61,Heaps_1978,Baayen}, 
which states that the growth of vocabulary with text length is sublinear (although the precise mathematical dependence has been debated
\cite{Font_Clos_Corral}). 


Two other laws, 
the law of word length and 
the so-called Zipf's law of abbreviation or brevity law,
are of particular interest in this work. 
As far as we know, and
in contrast to the Zipf's law of word frequency,
these two laws do not have non-linguistic counterparts.
The law of word length finds that the length of words (measured in number of letter tokens, for instance) is lognormally distributed \cite{Herdan58,Torre19}, 
whereas the brevity law determines that more frequent words tend to be shorter, whereas rarer words tend to be longer. 
This is usually quantified between a negative correlation between word frequency and word length \cite{Bentz_FerreriCancho}.

Very recently, Torre et al. \cite{Torre19} have 
parameterized
the dependence between mean frequency and length, 
obtaining (using a speech corpus) 
that the frequency averaged for fixed length 
decays exponentially with length.
This is in contrast with a result suggested by Herdan 
(to the best of our knowledge 
not directly supported by empirical analysis), 
who proposed a power-law decay, with exponent between 2 and 3
\cite{Herdan58}.
This result probably arose from an analogy with the 
word-frequency distribution derived by Simon \cite{Simon},
with an exponential tail that was neglected.

The purpose of our paper is to put these three important linguistic laws (Zipf's law of word frequency, the word-length law, and the brevity law) into a broader context. 
By means of considering word frequency and word length as two random variables associated to word types, we will see how the bivariate distribution of those two variables is the appropriate framework to describe the brevity-frequency phenomenon.
This leads us to several findings: 
(i) a gamma law for the word-length distribution,
in contrast to the previously proposed lognormal shape;
(ii) a well-defined functional form for the word-frequency distributions conditioned to fixed length, where a power-law decay with exponent $\alpha$ 
for the bulk frequencies becomes dominant;
(iii) a scaling law for those distributions,
apparent as a collapse of data under rescaling;
(iv) an approximate power-law decay of the characteristic scale of frequency
as a function of length, with exponent $\delta$;
and (v) a possible explanation for Zipf's law of word frequency
as arising from the mixture of conditional distributions of frequency at different
lengths, where Zipf's exponent is determined by the exponents $\alpha$ and $\delta$.

\section{Preliminary considerations}

Given a sample of natural language (a text, a fragment of speech, or a corpus, in general), any word type (i.e., each unique word) has an associated word length, which we measure in number of characters, and an associated word absolute frequency, which is the number of occurrences of the word type on the corpus under consideration 
(i.e., the number of tokens of the type). 
We denote these two random variables as $\ell$ and $n$, respectively.

Zipf's law of word frequency is written as a power-law relation between $f(n)$ and $n$ \cite{Moreno_Sanchez}, i.e.,
$$
f(n) \propto \frac 1 {n^\beta} 
\mbox { for } n \ge c,
$$
where $f(n)$ is the empirical probability mass function of the word frequency $n$, the symbol $\propto$ denotes proportionality, $\beta$ is the power-law exponent, and $c$ is a lower cut-off below which the law losses its validity 
(so, Zipf's law is a high-frequency phenomenon). 
The exponent $\beta$ takes values typically close to 2. 
When very large corpora are analyzed (made from many different texts an authors)
another (additional) power-law regime appears at smaller frequencies \cite{Ferrer2001a,Williams_Dodds}, 
$$
f(n) \propto \frac 1 {n^\alpha} 
\mbox { for } a \le n \le b,
$$
with $\alpha$ a new power law exponent smaller than $\beta$,
and $a$ and $b$ lower and upper cut-offs, respectively (with $a < b < c$).
This second power law is not identified with Zipf's law.

On the other side, the law of word lengths 
\cite{Herdan58}
proposes a lognormal distribution for the empirical probability mass function of word lengths, that is,
$$
f(\ell) \sim \mbox{LN}(\mu,\sigma^2),
$$
where LN denotes a lognormal distribution, whose associated normal distribution has mean $\mu$ and variance $\sigma^2$ 
(note that with the lognormal assumption 
it would seem that 
one is taking a continuous approximation for $f(\ell)$;
nevertheless, discreteness of $f(\ell)$ is still possible just redefining
the normalization constant).
The present paper challenges the lognormal law for $f(\ell)$.
Finally, the brevity law \cite{Bentz_FerreriCancho} can be summarized as
$$
\mbox{corr}(\ell,n) < 0,
$$
where $\mbox{corr}(\ell,n)$ is a correlation measure between $\ell$ and $n$, as for instance Pearson correlation, Spearman correlation, or Kendall correlation.

We claim that a more complete approach to the relationship between word length and word frequency can be obtained from the joint probability distribution $f(\ell,n)$ of both variables, together with the associated conditional distributions $f(n|\ell)$. 
To be more precise,
$f(\ell,n)$ is the joint probability mass function of type length and frequency, 
and $f(n|\ell)$ is the probability mass function of type frequency
conditioned to fixed length.
Naturally, the word-frequency distribution $f(n)$ 
and the word-length distribution $f(\ell)$ are just the two marginal distributions of $f(\ell,n)$. 

The relationships between these quantities are
$$
f(\ell) = \sum_{n=1}^\infty f(\ell,n),
$$
$$
f(n) = \sum_{\ell=1}^\infty f(\ell,n),
$$
$$
f(\ell,n) = f(n |\ell) f(\ell).
$$
Note that we will not use in this paper the equivalent relation $ f(\ell,n) = f(\ell | n) f(n)$, for sampling reasons ($n$ takes many more different values than $\ell$; so, for fixed values of $n$ one may find there is not enough statistics to obtain $f(\ell | n)$).
Obviously, all probability mass functions fulfil normalization,
$$
\sum_{\ell=1}^\infty \sum_{n=1}^\infty f(\ell,n)=
\sum_{n=1}^\infty f(n|\ell)=
\sum_{\ell=1}^\infty f(\ell)=
\sum_{n=1}^\infty f(n)=
1.
$$

We stress that, in our framework, 
each type yields one instance of the bivariate random variable $(\ell,n)$, 
in contrast to another equivalent approach for which it is each token that gives one instance of the (perhaps-different) random variables, see Ref. \cite{Corral_Cancho}. The use of each approach has important consequences for the formulation of Zipf's law, as it is well known \cite{Corral_Cancho}, 
and for the formulation of the word-length law 
(as it is not so well known \cite{Herdan58}).
Moreover, our bivariate framework is certainly different
to the that of Ref. \cite{Stephens_Bialek},
where the frequency was understood as a four-variate distribution
with the random variables taking 26 values from {\tt a} to {\tt z},
and also to the generalization in Ref. \cite{Corral_Muro}.

\section{Corpus and statistical methods}

We investigate the joint probability distribution of word-type length and frequency empirically, using
all English books in 
the recently presented Standardized Project Gutenberg Corpus
\cite{Gerlach_Font_Clos},
which comprises 
more than 40,000 books in English, 
with a total number of tokens equal to
2,016,391,406 
and a total number of types of
2,268,043. 
We disregard types 
with $n<10$ 
(relative frequency below $5\times 10^{-9}$)
and also those
not composed exclusively by the 26 usual letters from {\tt a} to {\tt z}
(previously, capital letters were transformed to lower-case).
This sub-corpus is further reduced by the elimination of
types with length above 20 characters,
in order to avoid spurious words
(among the eliminated types with $n\ge 10$ 
we only find three true English words:
{\tt 
incomprehensibilities, 
crystalloluminescence,}
and {\tt 
nitrosodimethylaniline}).
This reduces the numbers of tokens and types
respectively to 
2,010,440,020
and
391,529.
Thus,
all we need for our study is the list of all types (a dictionary)
including their absolute frequency $n$
and their length $\ell$ (measured in terms of number of characters).

Power-law distributions are fitted to the empirical data by using
the version for discrete random variables of 
the method for continuous distributions outlined in Ref. \cite{Peters_Deluca}
and developed in Refs. \cite{Corral_Deluca,Corral_Gonzalez},
which is based on maximum-likelihood estimation 
and the Kolmogorov-Smirnov goodness-of-fit test.
Acceptable (i.e., non-rejectable) fits require
$p-$values not below 0.20, 
which are computed with 1000 Monte-Carlo simulations.
Complete details in the discrete case are available in Refs.
\cite{Corral_Boleda,Moreno_Sanchez}.
This method is similar in spirit to the one by Clauset et al. \cite{Clauset},
but avoiding some of the important problems that the latter presents
\cite{Corral_nuclear,Voitalov_krioukov}.
Histograms are drawn to provide visual intuition for the 
shape of the empirical probability mass functions
and the adequacy of fits;
in the case of $f(n|\ell)$ and $f(n)$
we use logarithmic binning \cite{Corral_Deluca,Deluca_npg}.
Nevertheless, the computation of the fits does not make use of the 
graphical representation of the distributions.

On the other side,
the theory of scaling analysis, following Refs. \cite{Peters_Deluca,Corral_csf}, 
allows us to compare the shape of the conditional distributions $f(n|\ell)$ 
for different values of $\ell$.
This theory has revealed a very powerful tool in quantitative linguistics,
allowing 
in previous research
to show 
that the shape of the word-frequency distribution does
not change as a text increases its length
\cite{Font-Clos2013,Corral_Font_Clos_PRE17}.

\section{Results}
First, let us examine 
the raw data, looking at 
the scatter plot between frequency and length in Fig. 1, 
where each point is a word type represented by an associated value of $n$ and an associated value of $\ell$
(note that several or many types can overlap at the same point, 
if they share their values of $\ell$ and $n$,
as these are discrete variables). 
From the tendency of decreasing maximum $n$ with increasing $\ell$,
clearly visible in the plot, 
one could arrive to an erroneous version of the brevity law. 
Naturally, brevity would be apparent if the scatter plot were homogenously populated (i.e., if $f(\ell,n)$ would be uniform in the domain occupied by the points). 
But of course, this is not the case, as we will quantify later.
On the contrary, if $f(\ell,m)$ were the product of two independent exponentials, with $m=\ln n$,
the scatter plot would be rather similar to the real one (Fig. 1),
but the brevity law would not hold 
(because of the independence of $\ell$ and $m$, that is, of $\ell$ and $n$).
We will see that exponentials distributions play an important role here, 
but not in this way. 

\begin{figure}[!ht]
\begin{center}
\includegraphics[width=12cm]{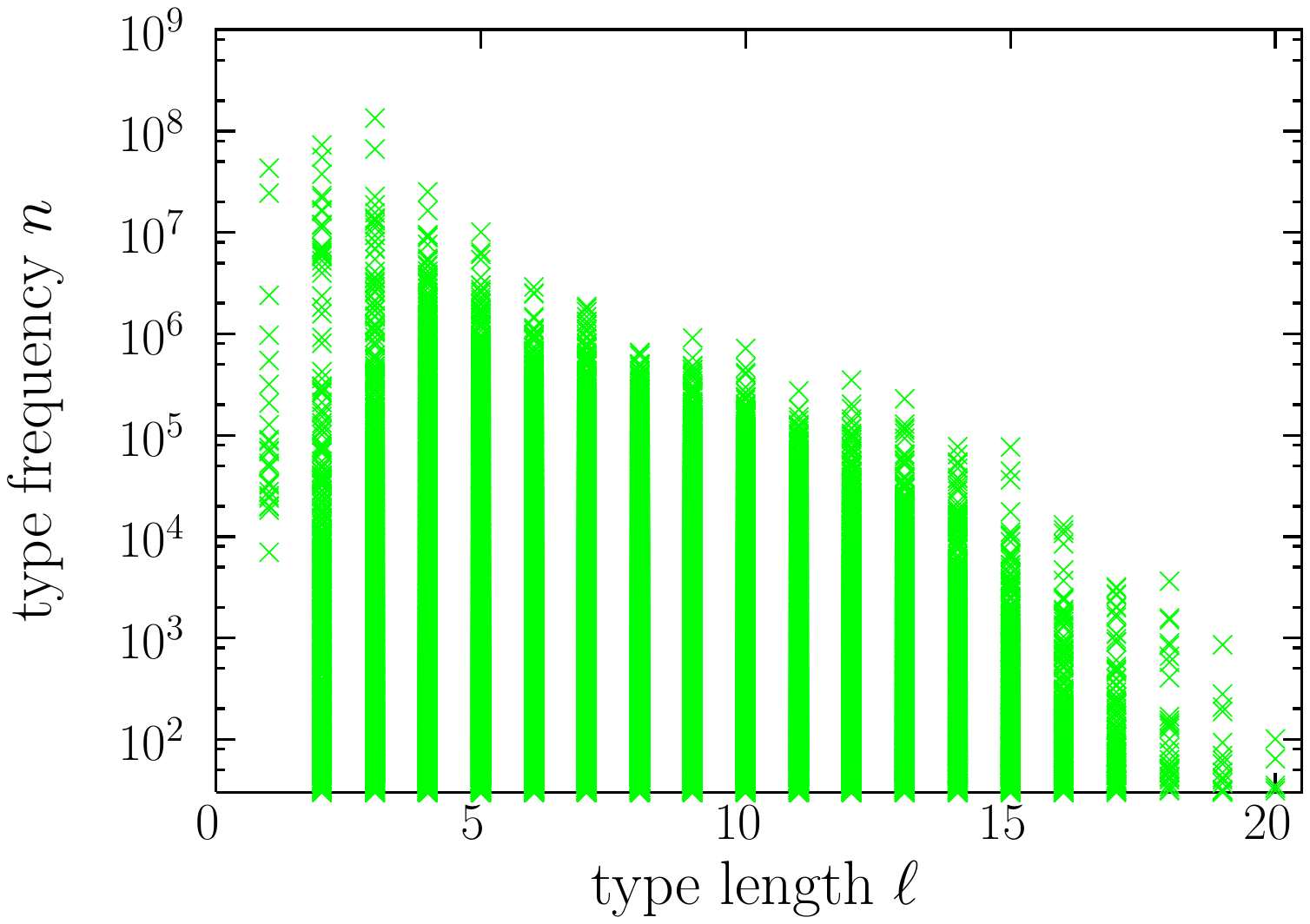}
\end{center}
\caption{
Illustration of the dataset by means of the
scatter plot between word-type frequency and length.
Frequencies below 30 are not shown.
}
\end{figure}

A more acceptable approach to the brevity-frequency phenomenon 
is to calculate the correlation between $\ell$ and $n$. 
For the Pearson correlation, our dataset yields $\mbox{corr}(\ell,n)=-0.023$, 
which, despite looking very small, is significantly different from zero,
with a $p-$value below 0.01 for 100 reshufflings of the frequency
(all the values obtained after reshuffling the frequencies keeping the lengths fixed
are between $-0.004$ and $0.006$).
If, instead, we calculate
the Pearson correlation between $\ell$ and the logarithm $m$ of the frequency
we get $\mbox{corr}(\ell, m)=-0.083$, 
again with a $p-$value below 0.01. 
Nevertheless, as neither the underlying joint distributions 
$f(\ell,n)$ or $f(\ell,m)$ 
resemble a Gaussian at all, nor the correlation seems to be linear 
(see Fig. 1), 
the meaning of the Pearson correlation is difficult to interpret.
We will see below that the analysis of the conditional distributions $f(n|\ell)$
provides more useful information.


\subsection{Marginal distributions}

\begin{figure}[!ht]
\begin{center}
\includegraphics[width=12cm]{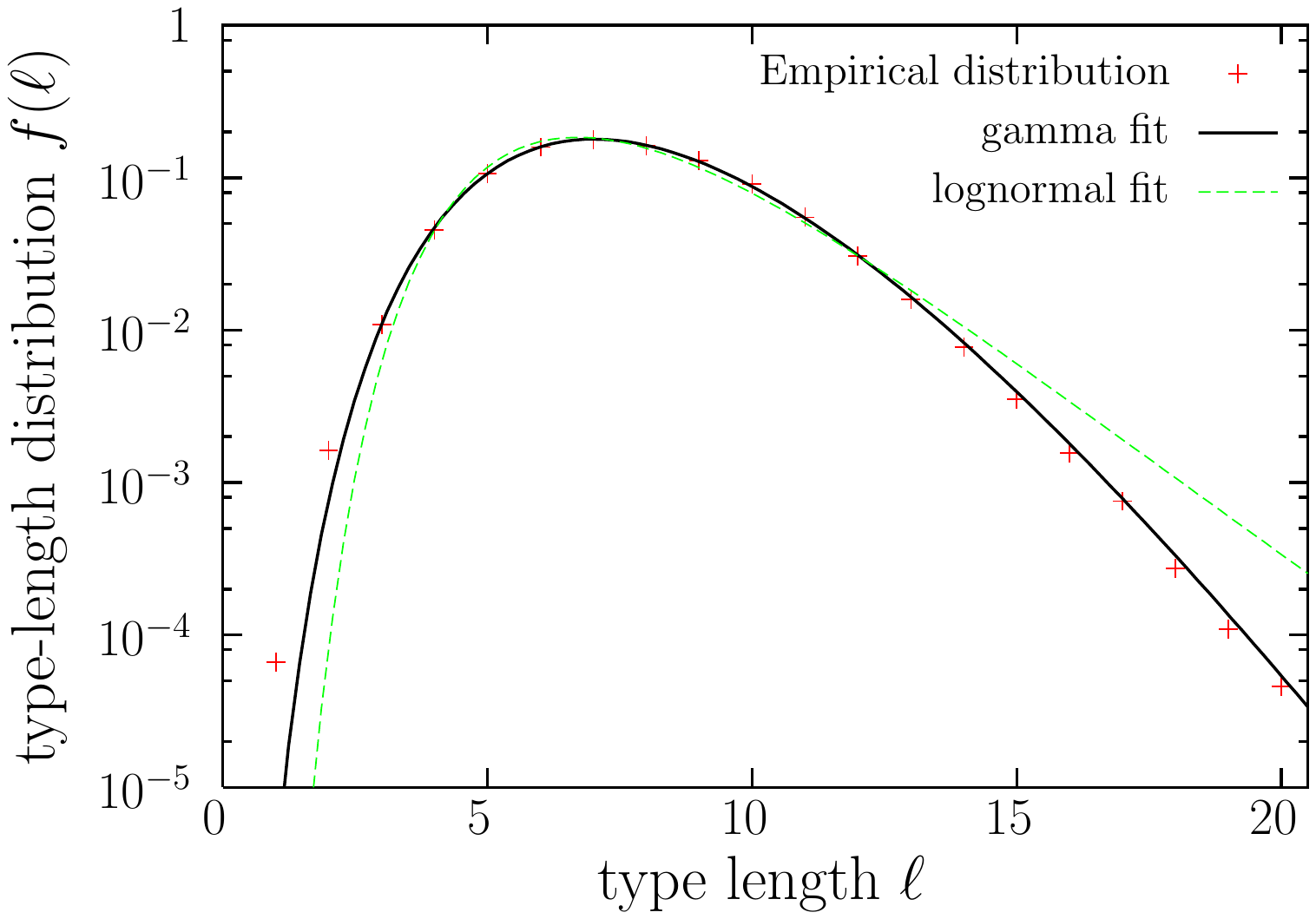}
\end{center}
\caption{
Probability mass function $f(\ell)$ of type length,
together with gamma and lognormal fits.
Note that the majority of types are those with lengths between 
4 and 13,
and that $f(\ell)$ is roughly constant between 5 and 10.
The superiority of the gamma fit is visually apparent,
and this is confirmed by loglikelihood equal to -872175.2
in front of the value -876535.1
for the lognormal
(a discrete gamma distribution slightly improves the fit, 
but the simple continuous case here is enough for our purposes).
The parameters resulting for the gamma fit are given in the text, 
and those for the lognormal are 
$\mu=1.9970\pm 0.0005$ and
$\sigma=0.3081\pm 0.0003$.
}
\label{Fig_length} 
\end{figure}

Let us now study the word-length distribution, $f(\ell)$, shown in Fig. \ref{Fig_length}. 
The distribution is clearly unimodal (with its maximum at $\ell=7$), 
and although it has been previously modelled as a lognormal \cite{Herdan58}, 
we get a nearly perfect fit using a gamma distribution,
\begin{equation}
f(\ell) =\frac \lambda {\Gamma(\gamma)} \left(\lambda \ell \right)^{\gamma-1} 
e^{-\lambda \ell},
\label{length_distribution}
\end{equation}
with shape parameter $\gamma=11.10 \pm 0.02$ and
inverted scale parameter $\lambda=1.439 \pm 0.003$
(where the uncertainty corresponds to one standard deviation,
and $\Gamma(\gamma)$ denotes the gamma function).
Notice then that, for large lengths, we would get an exponential decay
(asymptotically, strictly speaking).
However, there is an important difference between the lognormal distribution proposed in Ref. \cite{Torre19} and the gamma distribution found here, which is that the former case refers to the length of tokens, whereas in our case we deal with the length of types
(of course, length of tokens and length of types is the same length, 
but the relative number of tokens and types is different, depending on length).
This was already distinguished by Herdan \cite{Herdan58},
who used the terms occurrence distribution and dictionary distribution,
and proposed that both of them were lognormal.
In the caption of Fig. \ref{Fig_length} we provide the loglikelihoods of both
the gamma and lognormal fits, concluding that the gamma distribution yields a better fit for the ``dictionary distribution'' of word lengths.
The fit is specially good in the range $\ell > 2$.

Regarding the other marginal distribution, 
which is the word-frequency distribution $f(n)$ represented in Fig. 3, 
we get that, as expected, Zipf's law is fulfilled 
with $\beta=1.94\pm 0.03$ for $n \ge 1.9 \times 10^5$
(this is almost three orders of magnitude),
see Table \ref{tablecola}.
Another power-law regime in the bulk, as in Ref. 
\cite{Ferrer2001a}, is found to hold for one order of magnitude and a half (only),
from $a\simeq 400$ to $b\simeq 14000$,
with exponent $\alpha=1.41\pm 0.005$, see Table \ref{tablebulk}.
Note that although the truncated power law for the bulk part of the distribution
is much shorter than the one for the tail (1.5 orders of magnitude in front of almost 3), the former contains many more data (50000 in front of about 1000), 
see Tables \ref{tablecola} and \ref{tablebulk} for the precise figures.
Note also that the two power-law regimes for the frequency
translate into two exponential regimes for $m$ (the logarithm of $n$).

\begin{figure}[!ht]
\begin{center}
\includegraphics[width=12cm]{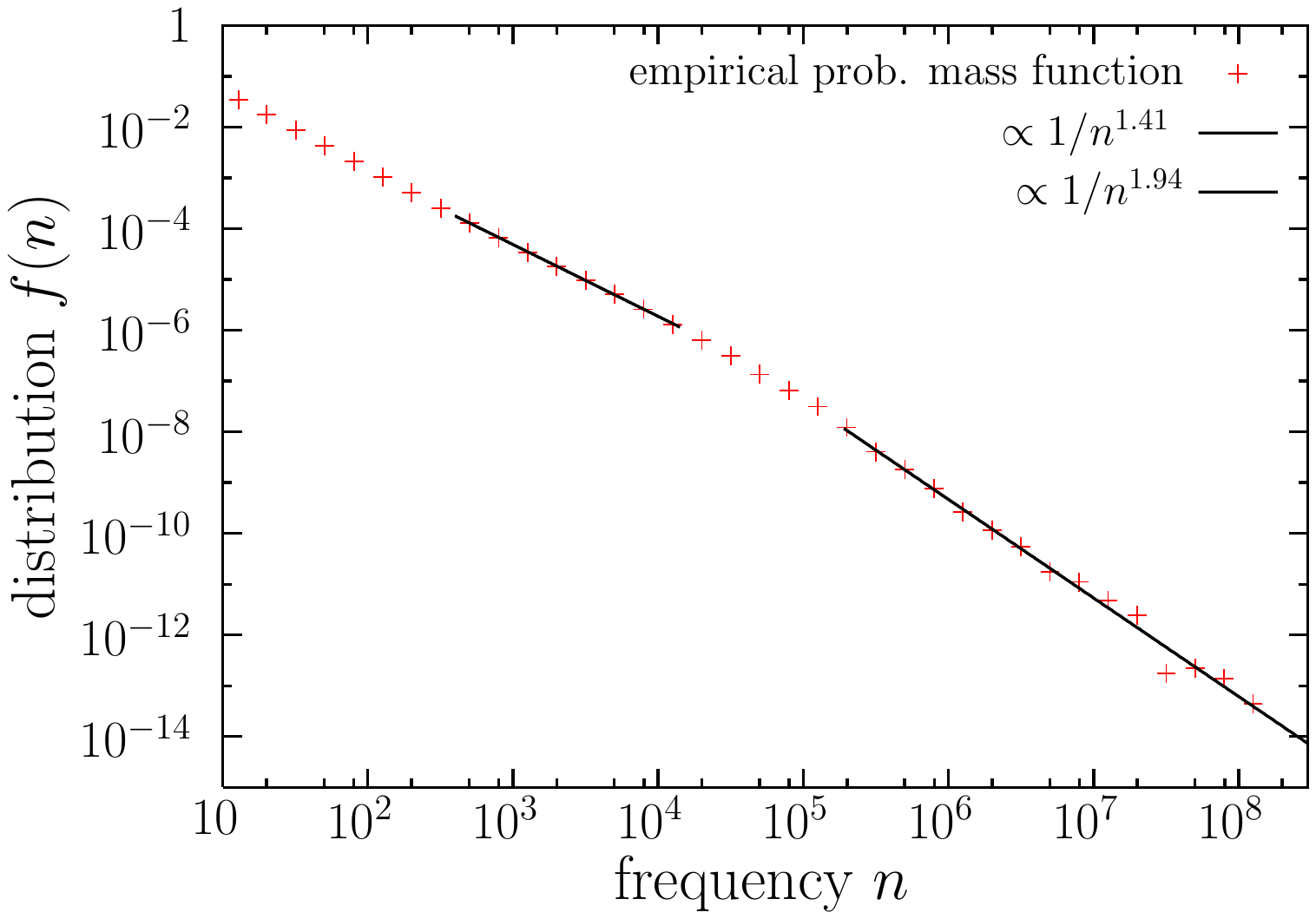}
\end{center}
\caption{
Probability mass function $f(n)$ of type frequency
(this is a marginal distribution with respect $f(\ell,n)$).
The results of the power-law fits are also shown.
The fit of a truncated continuous power law, maximizing number of data,
yields $\alpha=1.41$;
the fit of a untruncated discrete power law yields $\beta=1.94$.
}
\label{appen1}
\end{figure}
\subsection{Power laws and scaling law for the conditional distributions}

\begin{figure}[!ht]
\begin{center}
\includegraphics[width=12cm]{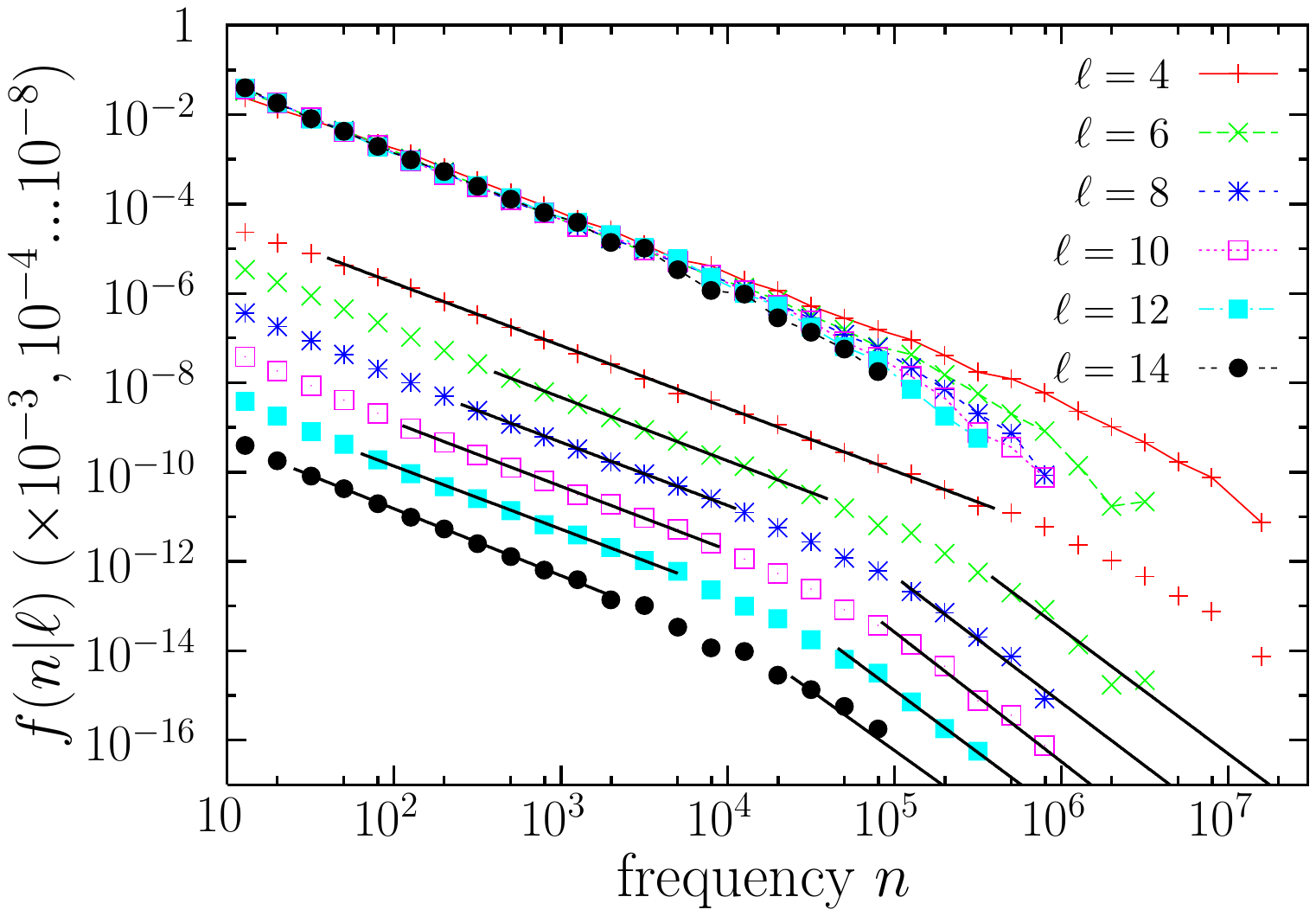}
\end{center}
\caption{
Probability mass functions $f(n|\ell)$ of frequency $n$
conditioned to fixed value of length $\ell$,
for several values of $\ell$. 
Distributions are shown twice: all together, 
and individually, displaced in the vertical axis by diverse factors $10^{-3}$, $10^{-4}$...
up to $10^{-8}$, for clarity sake of the power-law fits, represented by dark continuous lines.
}
\label{fig_cond} 
\end{figure}

\begin{figure}[!ht]
\begin{center}
\includegraphics[width=12cm]{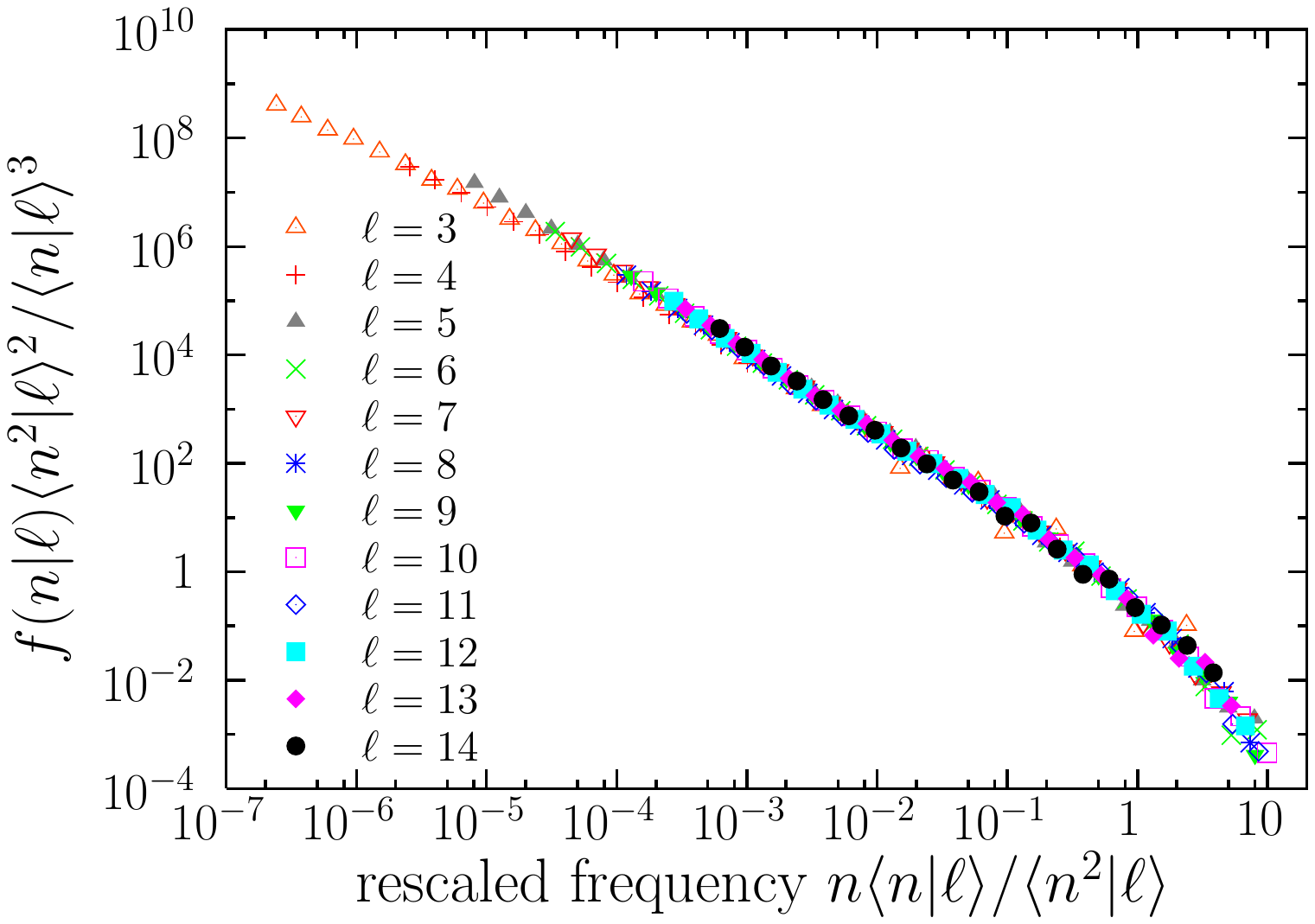}
\end{center}
\caption{
Word frequency probability mass functions $f(n|\ell)$ conditioned to fixed value of length rescaled by the ratio of powers of moments,
as a function as rescaled frequency,
for all values of length from 3 to 14.
The data collapse guarantees the fulfilment of a scaling law.
}
\label{fig_cond_sca} 
\end{figure}

As mentioned,
the conditional word-frequency distributions $f(n|\ell)$ are of substantial relevance. 
In Fig. \ref{fig_cond} we display some of those functions, 
turning out that $n$ is broadly distributed for each value of $\ell$
(roughly in the same qualitative way it happens without conditioning to the value of $\ell$).
Remarkably, the results of a scaling analysis
\cite{Peters_Deluca,Corral_csf}, depicted in Fig. \ref{fig_cond_sca}, 
show that
all the different $f(n|\ell)$ (for $3 \le \ell \le 14$) share a common shape, 
with a scale determined by a scale parameter in frequency.
Indeed, rescaling $n$ as $n \langle n | \ell \rangle / \langle n^2 | \ell \rangle$
and $f(n | \ell )$ as 
$f(n | \ell ) \langle n^2 | \ell \rangle^2 / \langle n | \ell \rangle^3 $,
where the first and second empirical moments,
$\langle n | \ell \rangle $ and
$\langle n^2 | \ell \rangle $, 
are also conditioned to the value of $\ell$,
we obtain an impressive data collapse, 
valid for about 7 orders of magnitude in $n$,
which allows us to write the scaling law 
$$
f(n|\ell) 
\simeq \frac{\langle n | \ell \rangle^3 }{\langle n^2 | \ell \rangle^2 } 
g\left(\frac{n\langle n | \ell \rangle }{\langle n^2 | \ell \rangle }\right)
\mbox{ for } 3 \le \ell \le 14,
$$
where the key point is that 
the scaling function $g(...)$ is the same function for any value of $\ell$.
For $\ell > 14$ the statistics is low and the fulfilment of the scaling law becomes uncertain.
Defining the scale parameter 
$\theta(\ell) = \langle n^2| \ell\rangle / \langle n|\ell\rangle$
we get alternative expressions for the same scaling law,
$$
f(n|\ell) \simeq 
\frac{\langle n | \ell \rangle}{\theta^2(\ell)} g\left(\frac{n}{\theta(\ell)}\right)
\propto
\frac{1}{\theta^\alpha(\ell)} g\left(\frac{n}{\theta(\ell)}\right)
\mbox{ for } 3 \le \ell \le 14,
$$
where constants of proportionality have been reabsorbed into $g$, and
the scale parameter has to be understood as proportional to a characteristic scale of the conditional distributions
(i.e., $\theta$ is the characteristic scale, up to a constant factor;
it is the relative change of $\theta$ what will be important for us).
The reason for the fulfilment of these relations is the power-law dependence
between the moments and the scale parameter when a scaling law holds, 
this power-law dependence is 
$\langle n |\ell \rangle \propto \theta^{2-\alpha}$
and 
$\langle n^2 |\ell \rangle \propto \theta^{3-\alpha}$
for $1<\alpha<2$,
see Refs. \cite{Peters_Deluca,Corral_csf}.

The data collapse also unveils more clearly the functional form of the scaling function $g$, 
allowing to fit its power-law shape in two different ranges.
The scaling function turns out to be 
compatible with a double power-law distribution,
i.e., a (long) power law for $n/\theta < 0.1$ with exponent $\alpha$ around 1.4 
and another (short) power law for $n/\theta > 1$ with exponent $\beta$ around 2.75; 
in one formula,
\begin{equation}
g(y) \propto 
\left\{\begin{array}{ll}
1/y^{1.4} & \mbox{ for } y\ll 1,\\
1/y^{2.75} & \mbox{ for } y> 1,\\
\end{array}
\right.
\label{doublepowerlaw}
\end{equation}
for $y=n/\theta$.
In other words, there is a (smooth) change of exponent 
(a change of log-log slope)
at a value of $n \simeq C \theta(\ell)$, with the proportionality constant $C$ 
taking some value in between 0.1 and 1
(as the transition from one regime to the other is smooth there is not a well
defined value of $C$ that separates both).
Fitting power laws to those ranges we get the results shown in Tables \ref{tablecola} and \ref{tablebulk}.
Note that $C\theta(\ell)$ can be understood as the characteristic scale
of $f(n|\ell)$ mentioned before, 
and can be also called a frequency crossover. 

Nevertheless, although the power-law regime for intermediate frequencies ($n<0.1\theta$) is very clear, 
the validity of the other power law (the one for large frequencies) is 
questionable,
in the sense that the power law provides an ``acceptable'' fit but other distributions could do the same good job, 
due to the limited range spanned by the tail (less than one order of magnitude).
Our main reason to fit a power law to the large-frequency regime is the comparison with Zipf's law ($\beta\simeq 2$),
and, as we see, the resulting value of $\beta$ for $f(n|\ell)$ 
turns out to be rather large
(the results of $\beta$ for all $f(n|\ell)$ turn out to be statistically compatible
with $\beta=2.75$).
In addition, we will show in the next subsection
that the high-frequency behavior of the conditional distributions (power law or not) has nothing to do with Zipf's law.

\begin{figure}[!ht]
\begin{center}
\includegraphics[width=12cm]{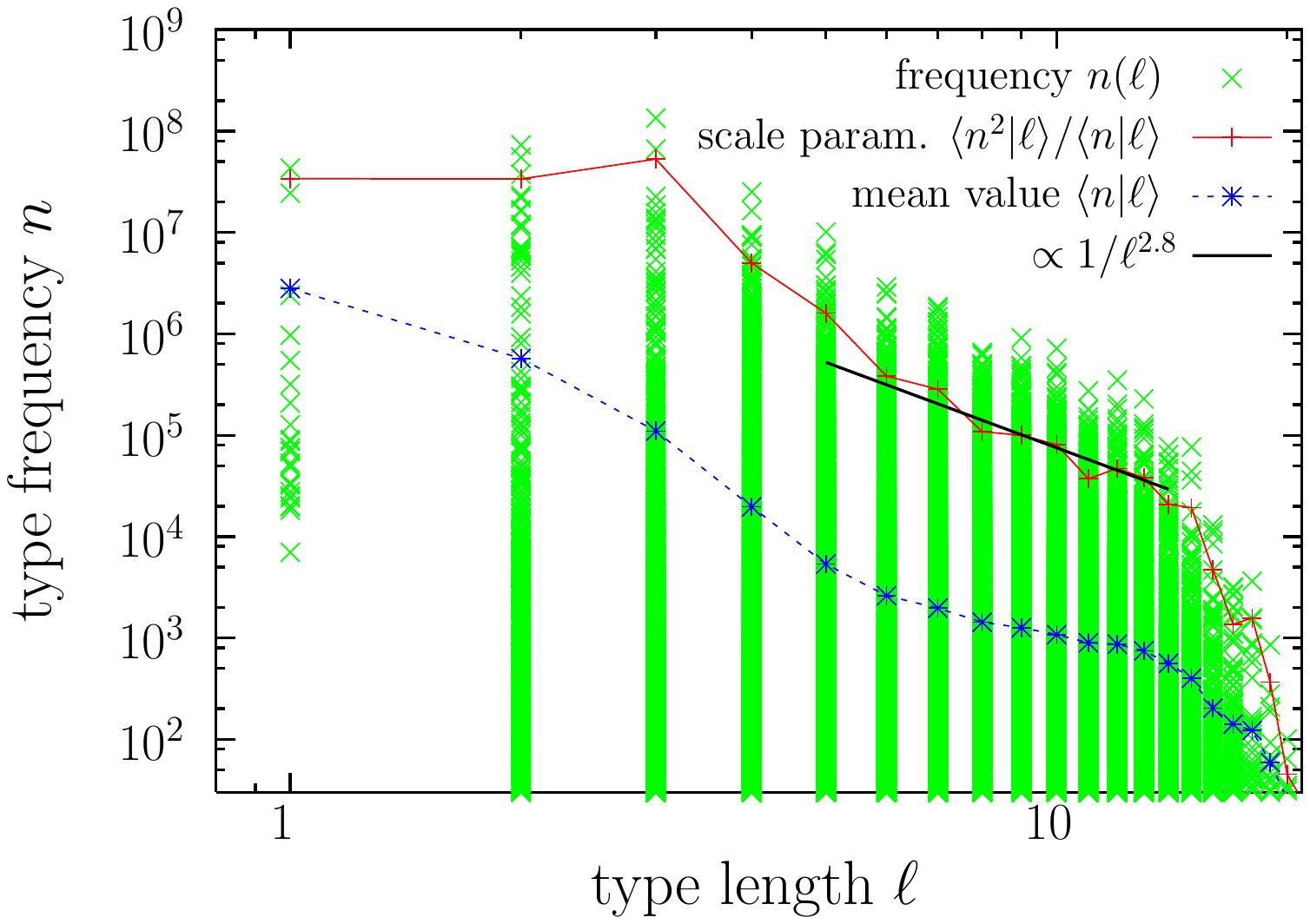}
\end{center}
\caption{
Scatter plot between word-type frequency and length (as in Fig. 1),
including also the estimated scale parameter $\theta=\langle n^2\ | \ell \rangle / \langle n\ | \ell \rangle$
and the conditional mean value $\langle n\ | \ell \rangle$.
A decaying power law with exponent 2.8, shown as a guide to the eye, 
is close to the empirical data for $ 6 \le \ell \le 13$.
Note that the horizontal axis is logarithmic, in contrast to Fig. 1.
Frequencies below 30 are not shown.
}
\end{figure}

\subsection{Brevity law and possible origin of Zipf's law}

Coming back to the scaling law, 
its fulfilment has an important consequence:
it is the scale parameter $\theta(\ell)$ 
and not the conditional mean $\langle n|\ell \rangle$
what sets the scale of the conditional distributions $f(n|\ell)$.
Figure 6 represents the brevity law in terms of the scale parameter 
as a function of $\ell$ (the conditional mean value is also shown, for comparison).
Note that Ref. \cite{Torre19} dealt with the conditional mean,
finding an exponential decay
$\langle n|\ell\rangle \propto 26^{-0.6\ell}$.
Using our corpus (which is certainly different), 
we find that such an exponential decay for the mean 
is valid in a range of $\ell$ between 1 and 5, roughly.
In contrast, the scale parameter $\theta$ shows an approximate
power-law decay from about $\ell=6$ to $15$,
with an exponent $\delta$ around 3 (or 2.8, to be more precise),
i.e., $$\theta(\ell) \propto 1/\ell^\delta$$
(note that Herdan assumed this exponent to be 2.4, 
with no clear empirical support \cite{Herdan58}).
Beyond $\ell=15$, the decay of $\theta(\ell)$ is much faster.
Nevertheless, these results are somewhat qualitative.

\begin{figure}[!ht]
\begin{center}
\includegraphics[width=12cm]{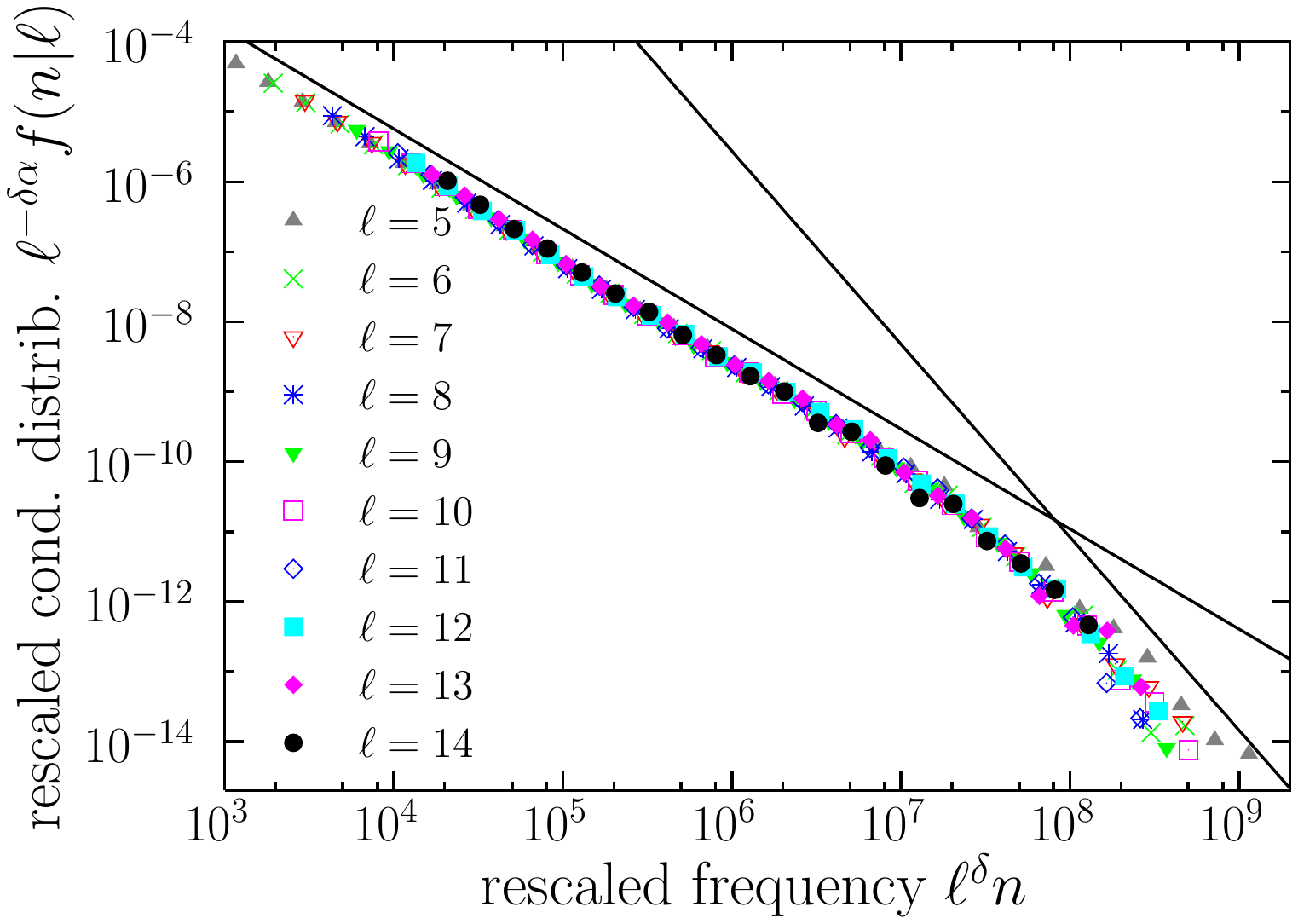}
\end{center}
\caption{
Same as Fig. 5, from $\ell=5$ to 14,
changing the scale factors from combination of powers
of moments ($\langle n | \ell \rangle$ and $\langle n | \ell \rangle$) 
to powers of length ($\ell^{-\delta}$ and $\ell^{\delta \alpha}$).
The collapse signals the fulfilment of a scaling law.
Two decreasing power laws with exponents 1.43 and 2.76 are shown
as straight lines, for comparison.
}
\label{fig_cond_sca2} 
\end{figure}

With these limitations, we could write a new version of the scaling law as
\begin{equation}
f(n|\ell) 
\simeq \ell^{\delta\alpha} g\left(\ell^\delta{n}\right)
\label{scalinglaw3}
\end{equation}
where the proportionality constant between $\theta$ and $\ell^\delta$
has been reabsorbed in the scaling function $g$.
The corresponding data collapse is shown in Fig. \ref{fig_cond_sca2},
for $5 \le \ell \le 14$.
Despite the rough approximation provided by the power-law decay of $\theta(\ell)$,
the data collapse in terms of scaling law (\ref{scalinglaw3}) is nearly excellent
for $\delta=2.8$.
This version of the scaling law provides a clean formulation of the brevity law:
the characteristic scale of the distribution of $n$ conditioned to the value of $\ell$
decays with increasing $\ell$ as $1/\ell^\delta$;
i.e., the larger $\ell$, the shorter the conditional distribution $f(n|\ell)$, 
quantified by the exponent $\delta$.

But, in addition to a new understanding of the brevity law, 
the scaling law in terms of $\ell$ provides, as a by-product, 
an empirical explanation of the origin of Zipf's law.
In the regime of $\ell$ in which the scaling law is approximately valid we
can obtain the distribution of frequency as a mixture of conditional distributions
(by the law of total probability), 
$$
f(n) = \int f(n|\ell) f(\ell) d\ell
$$
(where we take a continuous approximation, replacing sum over $\ell$ by integration;
this is essentially a mathematical rephrasing).
Substituting the scaling law and introducing the change of variables $x=\ell^\delta n$
$$
f(n)=\int \ell^{\delta\alpha} g\left(\ell^\delta{n}\right) f(\ell) d\ell
\propto \int \left(\frac x n\right)^\alpha g(x) \frac{x^{-1+1/\delta}}{n^{1/\delta}}dx
$$
where we also have taken advantage of the fact that in the region of interest, 
$f(\ell)$ can be considered (in a rough approximation) as constant.
From here it is clear that we get Zipf's law as
$$
f(n) \propto \frac 1 {n^{\alpha+\delta^{-1}}},
$$
i.e., Zipf's exponent can be obtained from the values of the intermediate-frequency power-law conditional exponent $\alpha$
and the brevity exponent $\delta$ as
$$
\beta_z=\alpha+\frac 1 \delta,
$$
where we have introduced a subscript $z$ in $\beta$
to stress that this is the $\beta$ exponent appearing in Zipf's law, 
corresponding to the marginal distribution $f(n)$, 
and to distinguish it from the one of the conditional distributions, 
that we may call $\beta_c$.
Note then that $\beta_c$ plays no role in the determination of $\beta_z$,
and, in fact, the scaling function does not need to have a power-law tail
in order that Zipf's law is obtained.
This sort of argument is similar to the one used in statistical seismology
\cite{Corral_calcutta}, but in that case the scaling law was elementary
(i.e., $\theta=\langle n|\ell \rangle$).

We can check the previous 
exponent relation using the empirical values of the exponent.
We do not have a unique measure of $\alpha$, but from Table \ref{tablebulk}
we see that its value for the different $f(n|\ell)$ is quite well defined.
Taking the harmonic mean between the values $4 \le \ell \le14$
we get $\bar \alpha=1.43$, which together with $\delta=2.8$ leads to
$\beta_z\simeq 1.79$, not far from the ideal Zipf's value $\beta_z=2$
and closer to the empirical value $\beta_z=1.94$.
The reason to calculate the harmonic mean of the exponents comes from the fact
that it is the maximum-likelihood outcome when untruncated power-law datasets
are put together \cite{Navas_pre2};
when the power laws are truncated, the result is closer to the untruncated case 
when the range $b/a$ is large.

\section{Conclusions}

Using a large corpus of English, 
we have seen how three important laws of quantitative linguistics, 
which are 
the type-length law, 
Zipf's law of word frequency, 
and the brevity law, can be put
into a unified framework just considering the joint distribution
of length and frequency.

Straightforwardly,
the marginals of the joint distribution provide 
both the type-length distribution
and the word-frequency distribution.
We reformulate the type-length law,
finding that the gamma distribution provides an
excellent fit of type lengths for values larger than 2, 
in contrast to the previously proposed 
lognormal distribution \cite{Herdan58}
(although some previous research was dealing not with type length
but with token length \cite{Torre19}).
For the distribution of word frequency we confirm the well-known Zipf's law, 
with an exponent $\beta_z=1.94$;
we also confirm 
the second intermediate power-law regime that emerges in large corpora
\cite{Ferrer2001a},
with an exponent $\alpha=1.4$.

The advantages of the perspective provided by considering
the length-frequency joint distribution become apparent
when dealing with the brevity phenomenon.
In concrete,
this property arises very clearly when looking at the distributions of frequency conditioned to fixed length. 
These show a well-defined shape, characterized by a power-law decay 
for intermediate frequencies followed by a faster decay, well modeled by a second power law, for larger frequencies.
The exponent $\alpha$ for the intermediate regime turns out to be the same
as the one for the usual (marginal) distribution of frequency, 
$\alpha\simeq 1.4$. However, the exponent for higher frequencies
$\beta_c$ turns out to be larger than 2 and unrelated to Zipf's law.

Scaling analysis reveals as a very powerful tool to explore and describe the brevity law. We observe that the conditional frequency distributions show scaling for different values of length, i.e., when the distributions are rescaled by a scale parameter (proportional to the characteristic scale of each distribution),
these distributions collapse into a unique curve, showing that they share a common shape (although at different scales).
The characteristic scale of the distributions turns out to be well described by the scale parameter (given by the ratio of moments $\langle n^2 |\ell\rangle/\langle n |\ell\rangle$), instead than by the mean value ($\langle n |\ell\rangle$).
This is the usual case when the distributions involved have a power-law shape
(with exponent $\alpha>1$) close to the origin \cite{Corral_csf}.
This also highlights the importance of looking at the whole distribution
and not to mean values when one is dealing with complex phenomena.

Going further, we obtain that the characteristic scale of the conditional frequency
distributions decays, approximately, as a power law of the type length, 
with exponent $\delta$,
which allows us to rewrite the scaling law in a form that is reminiscent to 
the one used in the theory of phase transitions and critical phenomena.
Despite that the power-law behavior for the characteristic scale of frequency
is rather rough, the derived scaling law shows an excellent agreement with the data.
Note that taking together the marginal length distribution, Eq.~(\ref{length_distribution}), 
and the scaling law for the conditional frequency distribution, Eq. (\ref{scalinglaw3}), 
we can write 
for the joint distribution
$$
f(\ell,n) \propto \lambda^\gamma \ell^{\delta\alpha+\gamma-1}
g(\ell^\delta n) e^{-\lambda \ell},
$$
with the scaling function $g(x)$ given by Eq. (\ref{doublepowerlaw}),
up to proportionality factors. 

Finally, the fulfilment of a scaling law of this form allows us to obtain a 
phenomenological (model free) explanation of Zipf's law as a mixture of the conditional distributions of frequencies. 
In contrast to some accepted explanations of Zipf's law, 
which put the origin of the law outside the linguistic realm
(such as Simon's model \cite{Simon}, where only the reinforced growth of the 
different types counts),
our approach shows that the origin of Zipf's law is fully linguistic, 
as it depends crucially on the length of the words:
at fixed length, each (conditional) frequency distribution
shows a scale-free (power-law) behavior, 
up to a characteristic frequency where the power law (with exponent $\alpha$)
breaks down. 
This breaking-down frequency depends on length through the exponent $\delta$.
The mixture of different power laws, with exponent $\alpha$,
cut at a scale governed by the exponent $\delta$ yields a
Zipf's exponent $\beta_z=\alpha+\delta^{-1}$.
Strictly speaking, our explanation of Zipf's law does not fully explain Zipf's law, 
but transfers the explanation to the existence of a power law with a smaller exponent ($\alpha\simeq 1.4$) as well as to the crossover frequency that
depends on length as $\ell^{-\delta}$.
Clearly, more research is necessary to explain the shape of the conditional distributions.

Although our results are obtained using a unique English corpus, 
we believe they are fully representative of this language, 
at least when large corpora are used.
Naturally, further investigations are needed to confirm this fact.
And of course, a necessary extension of our work 
is the use of corpora on other languages, 
to establish the universality of our results,
as done, e.g., in Ref. \cite{Bentz_FerreriCancho}.
The length of words is simply measured in number of characters,
but nothing precludes the use of number of phonemes or
time duration (in speech, as in Ref. \cite{Torre19}).
At the end, the goal of this kind of research is to pursue a
unified theory of linguistic laws, as proposed in Ref. \cite{Ferrer_cancho_compression}.
The venue shown in this paper seems to be a promising one.

\begin{table}[!ht]
\begin{center}
\caption{\label{tablecola} 
Results of the fitting of an discrete untruncated power law to the conditional distributions $f(n|\ell)$, denoted by a fixed $\ell$, 
and to the marginal distribution $f(n)$, denoted by the range $1 \le \ell \le 20$.
$N$ is total number of types, 
$n_{max}$ is the frequency of the most frequent type, 
$c$ is the lower cuf-off of the fit,
ordermag is $\log_{10} (n_{max}/c)$,
$N_c$ is the number of types with $n \ge c$,
$\beta$ is the resulting fitting exponent,
$\sigma_\beta$ is its standard deviation, 
and
$p$ is the $p-$value of the fit.
For the conditional distributions the possible fits are restricted to the range
$n > \langle n^2|\ell \rangle / \langle n|\ell \rangle$.
The fit proceeds by
sweeping 50 values of $c$ per order of magnitude
and using 1000 Monte-Carlo simulations for the calculation of $p$.
Of all the fits with $p \ge 0.20$ for a given $\ell$, the one with smaller $c$ is selected.
Outside the range $5 \le \ell \le 14$,
the number of types in the tail (below 10) is too low to yield a meaningful fit.
}
\smallskip
\begin{tabular}{ rrc ccr ccc}
$\ell$ & $N$ & $n_{max}$ & $c$ & ordermag & $N_{c}$ & $\beta \pm \sigma_\beta$ & $p$ \\ 
&& $(\times 10^5)$ & $(\times 10^5)$&&& \\ 
\hline
5& 41773 & 101. & 15.8 & 0.80 & 19 & 2.75$\pm$0.46 & 0.97\\
6 & 62277 & 29.0 & 3.80 & 0.88 & 60 & 2.79$\pm$0.24 & 0.31\\
7 & 69653 & 18.6 & 2.88 & 0.81 & 55 & 2.51$\pm$0.21 & 0.32\\
8 & 63574 & 6.55 & 1.10 & 0.78 & 133 & 2.82$\pm$0.17 & 0.25\\
9 & 50595 & 9.12 & 1.10 & 0.92 & 79 & 2.82$\pm$0.21 & 0.25\\
10 & 35679 & 7.16 & 0.83 & 0.93 & 69 & 2.90$\pm$0.24 & 0.75\\
11 & 21536 & 2.73 & 0.40 & 0.84 & 83 & 3.03$\pm$0.23 & 0.58\\
12 & 11973 & 3.49 & 0.46 & 0.88 & 34 & 2.78$\pm$0.33 & 0.65\\
13 & 6240 & 2.28 & 0.44 & 0.72 & 13 & 2.57$\pm$0.52 & 0.27\\
14 & 3035 & 0.77 & 0.24 & 0.51 & 12 & 2.67$\pm$0.56 & 0.22\\
\hline
$\le 20$ & 391529 & 1341 & 1.91 & 2.85 & 927 & 1.94$\pm$0.03 & 0.44\\
\hline
\end{tabular}
\par
\end{center}
\end{table}

\begin{table}[!ht]
\begin{center}
\caption{\label{tablebulk} 
Results of the fitting of a truncated power law to the conditional distributions $f(n|\ell)$, denoted by a fixed $\ell$, 
and to the marginal distribution $f(n)$, denoted by the range $1 \le \ell \le 20$.
$N$ is total number of types, 
$a$ and $b$ are the lower and upper cut-offs of the fit,
$N_{ab}$ is the number of types with $a\le n \le b$,
$\alpha$ is the resulting fitting exponent,
$\sigma_\alpha$ is its standard deviation, 
and
$p$ is the $p-$value of the fit.
The fit of a continuous power law is attempted in the range $n < 0.1 \langle n^2|\ell \rangle / \langle n|\ell \rangle$,
sweeping 20 values of $a$ and $b$ per order of magnitude
and using 1000 Monte-Carlo simulations for the calculation of $p$.
Of all the fits with $p \ge 0.20$ for a given $\ell$, the one with larger $b/a$ is selected, except for $f(n)$, where the largest $N_{ab}$ is used.
}
\smallskip
\begin{tabular}{ rrc ccr ccc}
$\ell$ & $N$ & $a$ & $b$ & ordermag & $N_{ab}$ & $\alpha \pm \sigma_\alpha$ & $p$ \\ 
\hline
1 & 26 & .126E+05 & .251E+07 & 2.30 & 23 & 1.391$\pm$0.155 & 0.24\\
2 & 636 & .200E+04 & .251E+07 & 3.10 & 188 & 1.486$\pm$0.045 & 0.24\\
3 & 4282 & .794E+03 & .447E+07 & 3.75 & 1171 & 1.428$\pm$0.016 & 0.30\\
4 & 17790 & .398E+02 & .398E+06 & 4.00 & 10618 & 1.402$\pm$0.005 & 0.20\\
5 & 41773 & .562E+03 & .178E+06 & 2.50 & 5747 & 1.426$\pm$0.009 & 0.37\\
6 & 62277 & .398E+03 & .398E+05 & 2.00 & 8681 & 1.421$\pm$0.009 & 0.27\\
7 & 69653 & .200E+03 & .282E+05 & 2.15 & 13392 & 1.449$\pm$0.007 & 0.25\\
8 & 63574 & .251E+03 & .112E+05 & 1.65 & 9849 & 1.417$\pm$0.010 & 0.41\\
9 & 50595 & .200E+03 & .100E+05 & 1.70 & 8850 & 1.400$\pm$0.010 & 0.25\\
10 & 35679 & .112E+03 & .891E+04 & 1.90 & 8454 & 1.428$\pm$0.010 & 0.21\\
11 & 21536 & .562E+02 & .141E+04 & 1.40 & 6227 & 1.469$\pm$0.015 & 0.22\\
12 & 11973 & .631E+02 & .501E+04 & 1.90 & 3866 & 1.411$\pm$0.013 & 0.51\\
13 & 6240 & .562E+02 & .398E+04 & 1.85 & 2144 & 1.396$\pm$0.019 & 0.90\\
14 & 3035 & .251E+02 & .224E+04 & 1.95 & 1567 & 1.496$\pm$0.022 & 0.27\\
15 & 1384 & .224E+02 & .200E+04 & 1.95 & 777 & 1.488$\pm$0.031 & 0.59\\
16 & 612 & .282E+02 & .447E+03 & 1.20 & 256 & 1.569$\pm$0.082 & 0.22\\
17 & 296 & .126E+02 & .141E+03 & 1.05 & 205 & 1.784$\pm$0.110 & 0.24\\
18 & 107 & .112E+02 & .158E+03 & 1.15 & 79 & 2.008$\pm$0.172 & 0.28\\
\hline
$\le 20$ & 391529 & .398E+03 & .141E+05 & 1.55 & 51972 & 1.413$\pm$0.005 & 0.21\\
\hline
\end{tabular}
\par
\end{center}
\end{table}

\section{Acknowledgements}

We are indebted to F. Font-Clos for providing the valuable corpus
released in Ref. \cite{Gerlach_Font_Clos}.
Our interest in the brevity law arised from our interaction with R. Ferrer-i-Cancho, 
in particular from the reading of Refs. \cite{ferrericancho2015compression,Ferrer_cancho_compression}. 
Support from projects
FIS2012-31324, FIS2015-71851-P, 
PGC-FIS2018-099629-B-I00,
Mar\'{\i}a de Maeztu Program 
MDM-2014-0445,
from Spanish MINECO
and the Collaborative Mathematics Project from La Caixa Foundation
(I.S.) is acknowledged.


\begin{thebibliography}{10}

\bibitem{Hdez_FCancho_book}
T.~Hern\'andez and R.~{Ferrer i Cancho}.
\newblock {\em Ling\"u\'{\i}stica Cuantitativa}.
\newblock El Pa\'{\i}s Ediciones, Madrid, 2019.

\bibitem{Zipf_1949}
G.~K. Zipf.
\newblock {\em Human Behavior and the Principle of Least Effort}.
\newblock Addison-Wesley, 1949.

\bibitem{Baayen}
H.~Baayen.
\newblock {\em Word Frequency Distributions}.
\newblock Kluwer, Dordrecht, 2001.

\bibitem{Baroni2009}
M.~Baroni.
\newblock Distributions in text.
\newblock In A.~L\"udeling and M.~Kyt\"o, editors, {\em Corpus linguistics: An
  international handbook, Volume 2}, pages 803--821. Mouton de Gruyter, Berlin,
  2009.

\bibitem{Zanette_book}
D.~Zanette.
\newblock Statistical patterns in written language.
\newblock {\em arXiv}, 1412.3336v1, 2014.

\bibitem{Piantadosi}
S.~T. Piantadosi.
\newblock Zipf's law in natural language: a critical review and future
  directions.
\newblock {\em Psychon. Bull. Rev.}, 21:1112--1130, 2014.

\bibitem{Moreno_Sanchez}
I.~Moreno-S\'anchez, F.~Font-Clos, and A.~Corral.
\newblock Large-scale analysis of {Zipf}'s law in {English} texts.
\newblock {\em PLoS ONE}, 11(1):e0147073, 2016.

\bibitem{Corral_Cancho}
A.~Corral, I.~Serra, and R.~Ferrer{-i-}Cancho.
\newblock The distinct flavors of {Zipf}'s law in the rank-size and in the
  size-distribution representations, and its maximum-likelihood fitting.
\newblock {\em {arXiv}}, 1908:01398, 2019.

\bibitem{Mandelbrot61}
B.~Mandelbrot.
\newblock {On the theory of word frequencies and on related {Markovian} models
  of discourse}.
\newblock In R.~Jakobson, editor, {\em Structure of Language and its
  Mathematical Aspects}, pages 190--219. American Mathematical Society,
  Providence, RI, 1961.

\bibitem{Heaps_1978}
H.~S. Heaps.
\newblock {\em Information retrieval: computational and theoretical aspects}.
\newblock Academic Press, 1978.

\bibitem{Font_Clos_Corral}
F.~Font-Clos and A.~Corral.
\newblock Log-log convexity of type-token growth in {Zipf}'s systems.
\newblock {\em Phys. Rev. Lett.}, 114:238701, 2015.

\bibitem{Herdan58}
G.~Herdan.
\newblock {The Relation Between the Dictionary Distribution and the Occurrence
  Distribution of Word Length and its Importance for the Study of Quantitative
  Linguistics}.
\newblock {\em Biometrika}, 45(1--2):222--228, 06 1958.

\bibitem{Torre19}
I.~G. Torre, B.~Luque, L.~Lacasa, C.~T. Kello, and A.~Hern\'andez-Fern\'andez.
\newblock On the physical origin of linguistic laws and lognormality in speech.
\newblock {\em Royal Soc. Open Sci.}, 6(8):191023, 2019.

\bibitem{Bentz_FerreriCancho}
C.~Bentz and R.~{Ferrer-i-Cancho}.
\newblock Zipf's law of abbreviation as a language universal.
\newblock In C.~Bentz, G.~J\"ager, and I.~Yanovich, editors, {\em Proceedings
  of the Leiden Workshop on Capturing Phylogenetic Algorithms for Linguistics}.
  University of T\"ubingen, 2016.

\bibitem{Simon}
H.~A. Simon.
\newblock On a class of skew distribution functions.
\newblock {\em Biomet.}, 42:425--440, 1955.

\bibitem{Ferrer2001a}
R.~{Ferrer i Cancho} and R.~V. Sol\'e.
\newblock Two regimes in the frequency of words and the origin of complex
  lexicons: {Zipf's} law revisited.
\newblock {\em J. Quant. Linguist.}, 8(3):165--173, 2001.

\bibitem{Williams_Dodds}
J.~R. Williams, J.~P. Bagrow, C.~M. Danforth, and P.~S. Dodds.
\newblock Text mixing shapes the anatomy of rank-frequency distributions.
\newblock {\em Phys. Rev. E}, 91:052811, 2015.

\bibitem{Stephens_Bialek}
G.~J. Stephens and W.~Bialek.
\newblock Statistical mechanics of letters in words.
\newblock {\em Phys. Rev. E}, 81:066119, 2010.

\bibitem{Corral_Muro}
A.~Corral and M.~{Garc\'ia del Muro}.
\newblock From {Boltzmann} to {Zipf} through {Shannon} and {Jaynes}.
\newblock {\em arXiv}, 1912.03570, 2019.

\bibitem{Gerlach_Font_Clos}
M.~Gerlach and F.~Font{-}Clos.
\newblock A standardized {Project Gutenberg} corpus for statistical analysis of
  natural language and quantitative linguistics.
\newblock {\em arXiv}, 1812.08092, 2018.

\bibitem{Peters_Deluca}
O.~Peters, A.~Deluca, A.~Corral, J.~D. Neelin, and C.~E. Holloway.
\newblock Universality of rain event size distributions.
\newblock {\em J. Stat. Mech.}, P11030, 2010.

\bibitem{Corral_Deluca}
A.~Deluca and A.~Corral.
\newblock Fitting and goodness-of-fit test of non-truncated and truncated
  power-law distributions.
\newblock {\em Acta Geophys.}, 61:1351--1394, 2013.

\bibitem{Corral_Gonzalez}
A.~Corral and A.~Gonz\'alez.
\newblock Power law distributions in geoscience revisited.
\newblock {\em Earth Space Sci.}, 6(5):673--697, 2019.

\bibitem{Corral_Boleda}
A.~Corral, G.~Boleda, and R.~{Ferrer-i-Cancho}.
\newblock Zipf's law for word frequencies: Word forms versus lemmas in long
  texts.
\newblock {\em PLoS ONE}, 10(7):e0129031, 2015.

\bibitem{Clauset}
A.~Clauset, C.~R. Shalizi, and M.~E.~J. Newman.
\newblock Power-law distributions in empirical data.
\newblock {\em SIAM Rev.}, 51:661--703, 2009.

\bibitem{Corral_nuclear}
A.~Corral, F.~Font, and J.~Camacho.
\newblock Non-characteristic half-lives in radioactive decay.
\newblock {\em Phys. Rev. E}, 83:066103, 2011.

\bibitem{Voitalov_krioukov}
I.~{Voitalov}, P.~{van der Hoorn}, R.~{van der Hofstad}, and D.~{Krioukov}.
\newblock Scale-free networks well done.
\newblock {\em Phys. Rev. Research}, 1:033034, 2019.

\bibitem{Deluca_npg}
A.~Deluca and A.~Corral.
\newblock Scale invariant events and dry spells for medium-resolution local
  rain data.
\newblock {\em Nonlinear Proc. Geophys.}, 21:555--567, 2014.

\bibitem{Corral_csf}
A.~Corral.
\newblock Scaling in the timing of extreme events.
\newblock {\em Chaos. Solit. Fract.}, 74:99--112, 2015.

\bibitem{Font-Clos2013}
F.~Font-Clos, G.~Boleda, and A.~Corral.
\newblock A scaling law beyond {Zipf}'s law and its relation to {Heaps}' law.
\newblock {\em New J. Phys.}, 15:093033, 2013.

\bibitem{Corral_Font_Clos_PRE17}
A.~Corral and F.~Font-Clos.
\newblock Dependence of exponents on text length versus finite-size scaling for
  word-frequency distributions.
\newblock {\em Phys. Rev. E}, 96:022318, 2017.

\bibitem{Corral_calcutta}
A.~Corral.
\newblock Statistical features of earthquake temporal occurrence.
\newblock In P.~Bhattacharyya and B.~K. Chakrabarti, editors, {\em Modelling
  Critical and Catastrophic Phenomena in Geoscience}, Lecture Notes in Physics,
  705, pages 191--221. Springer, Berlin, 2007.

\bibitem{Navas_pre2}
V.~Navas-Portella, I.~Serra, A.~Corral, and E.~Vives.
\newblock Increasing power-law range in avalanche amplitude and energy
  distributions.
\newblock {\em Phys. Rev. E}, 97:022134, 2018.

\bibitem{Ferrer_cancho_compression}
R.~{Ferrer-i-Cancho}.
\newblock Compression and the origins of {Zipf's} law for word frequencies.
\newblock {\em Complexity}, 21:409--411, 2016.

\bibitem{ferrericancho2015compression}
R.~{Ferrer-i-Cancho}, C.~Bentz, and C.~Seguin.
\newblock Compression and the origins of {Zipf's} law of abbreviation.
\newblock {\em arXiv}, 1504.04884, 2015.

\end{thebibliography}

\end{document}